\documentclass[aip,rsi,reprint]{revtex4-1}
\usepackage{graphicx}
\begin{document}

\title{A scripted control system for autonomous hardware-timed experiments}

\author{P.~T.~Starkey}
\thanks{P.~T.~Starkey and C.~J.~Billington contributed equally to this work.}
\email[Correspondence should be addressed to P. T. Starkey, electronic mail: ]{philip.starkey@monash.edu}
\author{C.~J.~Billington}
\thanks{P.~T.~Starkey and C.~J.~Billington contributed equally to this work.}
\email[Correspondence should be addressed to P. T. Starkey, electronic mail: ]{philip.starkey@monash.edu}
\author{S.~P.~Johnstone}
\author{M.~Jasperse}
\author{K.~Helmerson}
\author{L.~D.~Turner}
\author{R.~P.~Anderson}
\noaffiliation
\affiliation{School of Physics, Monash University, Victoria 3800, Australia.}

\date{\today}

\begin{abstract}
We present the \textit{labscript suite}, an open-source experiment control system for automating shot-based experiments and their analysis.
Experiments are composed as Python code, which is used to produce low-level hardware instructions.
They are queued up and executed on the hardware in real time, synchronized by a pseudoclock.
Experiment parameters are manipulated graphically, and analysis routines are run as new data are acquired.
With this system, we can easily automate exploration of parameter spaces, including closed-loop optimization.
\end{abstract}

\maketitle

\section{Introduction}
Modern experiments in quantum science demand flexible, autonomous control of heterogeneous hardware.
Many such experiment are \textit{shot}-based: a single experiment shot comprises analog, digital, and radiofrequency (rf) outputs operating under precise timing, as well as synchronized camera exposures and voltage measurements.
Bose--Einstein condensation (BEC) experiments\cite{[{See, e.g., }][{ and references within.}]weidemuller_cold_2009}, for example, require a timing resolution down to a few hundred nanoseconds, and may last for up to a minute.
Output must, therefore, be hardware timed, requiring devices be programmed with instructions in advance of an experiment shot.
Most measurements of interest require numerous shots, to build up statistics, or to observe the response of the system to varying parameters.
Such repetition is common to experiments employing cold quantum gases or trapped ions for precision metrology\cite{[{See, e.g., }][]robins_atom_2013,*[][{ and references within.}]cronin_optics_2009}, quantum computation\cite{[{See, e.g., }][]negretti_quantum_2011,*[][{ and references within.}]ladd_quantum_2010}, and quantum simulation\cite{[{See, e.g., }][]bloch_quantum_2012,*[][{ and references within.}]blatt_quantum_2012}.

Individual shots are typically complex, requiring the coordination of many devices.
This coordination is the role of a \textit{control system}.
A good control system should automate the programming of devices based on a high-level description of the experiment logic\cite{varoquaux_agile_2008}.
It should handle the repetition of shots and automated variation of experiment parameters, the increasingly complex demands of which cannot be rapidly, robustly, and continuously met by human operators.
It should automate analysis, leading to the prospect of closed-loop control: the results of analysis influencing subsequent experiment shots.
Applications of such closed-loop control include autonomous algorithmic optimization of parameters, and automatic recalibration in response to environmental drifts. 

Most existing control systems take one of two approaches for providing a human interface to programming hardware.
One is text-based, in which experiments are written using a general purpose programming language\cite{gaskell_open-source_2009}. 
In the other, experiments are instead described graphically using a custom user interface\cite{anderson_nonequilibrium_2010,beeler_matthew_disordered_2011,altin_role_2012,stoferle_exploring_2005,keshet_distributed_2013}.
The text-based approach natively offers the advantages of a programming language, particularly control-flow tools such as conditional statements, loops, and functions.
Its disadvantage is that frequently varied settings and parameters may be hidden in hundreds of lines of code.
Conversely, the graphical-user-interface (GUI) approach makes experiment parameters more accessible to the user, but features providing for complex experiment logic must be anticipated and implemented specifically\cite{keshet_distributed_2013,stoferle_exploring_2005}.

The two approaches need not be mutually exclusive: by separating experiment parameters from experiment logic, parameters can be manipulated graphically and logic textually\cite{owen_fast_2004,meyrath_laboratory_2012}.
We contend that by using a high-level programming language with appropriate hardware abstraction, text-based control can be more comprehensible to a newcomer than an equivalent graphical representation of hardware instructions.

We present the \textit{labscript suite} which utilizes a hybrid text-and-GUI approach for control and builds on previous work by addressing the need for autonomous control, analysis, and optimization.
Hardware control is abstracted, providing an identical software interface to devices of a common type.
Graphical interfaces are dynamically generated based on the current hardware set in use.
Analysis is an integral part of the control system, with user-written analysis routines run automatically on new data.
Finally, analysis results can modify subsequent experiment shots, closing the feedback loop on analysis and control.

\begin{figure*}%
\includegraphics{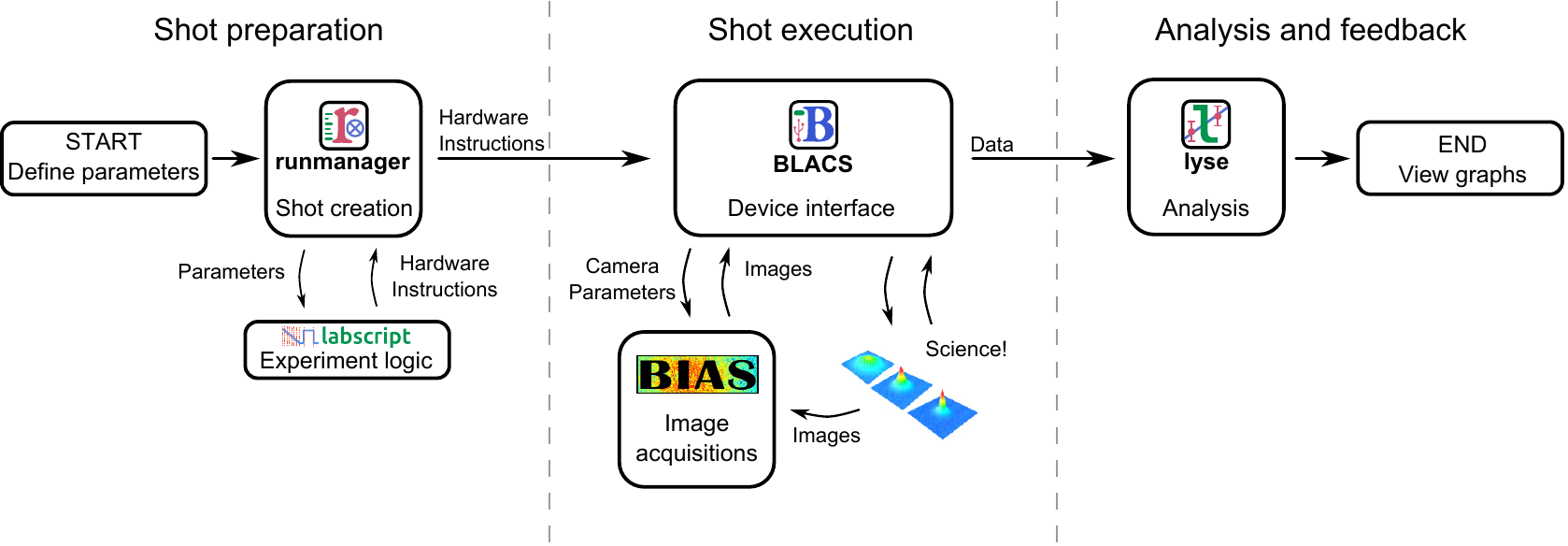}%
\caption{Each experiment shot comprises three stages: preparation, execution, and analysis.
Arrows indicate how the HDF file for an experiment shot passes between software components of the labscript suite.
Only the shot execution stage is coupled to hardware timing, allowing new shots to be created and queued while others are running.
Similarly, analysis can be performed on executed shots at any time.
}

\label{flow_chart}%
\end{figure*}

\section{An overview of the labscript suite} 
The labscript suite comprises several programs, each performing one main function; the flow of data between programs is shown in Fig.~\ref{flow_chart}.
Each experiment shot is associated with a single file: each program writes to and reads from this file as required before passing it on to the next program.
Programs may be run on separate computers, communicating over the network using the ZeroMQ messaging library\cite{[][{; see also ``\O MQ: the intelligent transport layer,'' \url{http://www.zeromq.org/.}}]hintjens_code_2013}, exchanging references to the experiment file.

We use the Hierarchical Data Format (HDF version 5)\cite{the_hdf_group_hdf5_2012} which provides cross-platform storage of large scientific datasets.
Exploiting the extensibility of HDF, each file is a complete description of the experiment shot.
The HDF file begins life containing only experiment parameters.
As it is passed between components of the labscript suite, the file grows to contain the hardware instructions, acquired, data and analysis results.
Metadata is also stored including user-written scripts and version control information.
This maintains a comprehensive record of the experiment shot for post-hoc analysis, reproducibility, and publication preparation.

Attempts to standardize laboratory device programming have largely failed, with only a minority of devices conforming to standards such as SCPI (Standard Commands for
Programmable Instruments).\cite{scpi}
This calls for abstraction to shield the user from low-level interaction.
We have created a software library for Python\cite{rossum_python,hughes_real_2010}, \texttt{labscript} (Sec.~\ref{metalanguage}), which provides a common interface for commanding output and measurements from devices.
The user writes the experiment logic in Python, and \texttt{labscript}
generates the required hardware instructions, including a clocking signal for timing (Sec.~\ref{pseudoclock}).

The labscript suite separates experiment logic (written in Python) from experiment parameters, which are manipulated in a GUI.
The GUI, \texttt{runmanager} (Sec.~\ref{runmanager}), creates the HDF file for the experiment shot and stores the parameters within.
If a parameter is a list of values, rather than a single value, \texttt{runmanager} creates an HDF file (a prospective shot) for each value.
If lists are entered for more than one parameter, \texttt{runmanager} creates a file for each point in the resulting parameter space.

For each shot, \texttt{labscript} inserts the parameters from the HDF file into the experiment logic, compiles hardware instructions for each device, and writes them to the same file.
\texttt{runmanager} sends the compiled HDF files to \texttt{BLACS} (Sec.~\ref{experiment_execution}) which places them in a queue.
\texttt{BLACS} interfaces with hardware devices either directly, or via secondary control programs such as \texttt{BIAS} (Sec.~\ref{BIAS}).
\texttt{BLACS} programs the hardware and triggers the experiment shot to begin.
The experiment then proceeds under hardware-timed control.

Once the experiment shot has finished, acquired data such as voltage time-series and images are added to the HDF file.
\texttt{BLACS} then passes the file to a dedicated analysis system, \texttt{lyse} (Sec.~\ref{lyse}).
\texttt{lyse} coordinates the execution of analysis routines, which are Python scripts written by the user.
These scripts may analyze individual shots or a sequence of shots as a whole.
This facilitates autonomous analysis of results from parameter space scans, as experiment shots are completed.

The \texttt{labscript} software library can be applied to automatically generate shots based on the results of analysis.
We have used this to implement a closed-loop optimization system, \texttt{mise} (Sec.~\ref{mise}).

\begin{figure*}%
\includegraphics{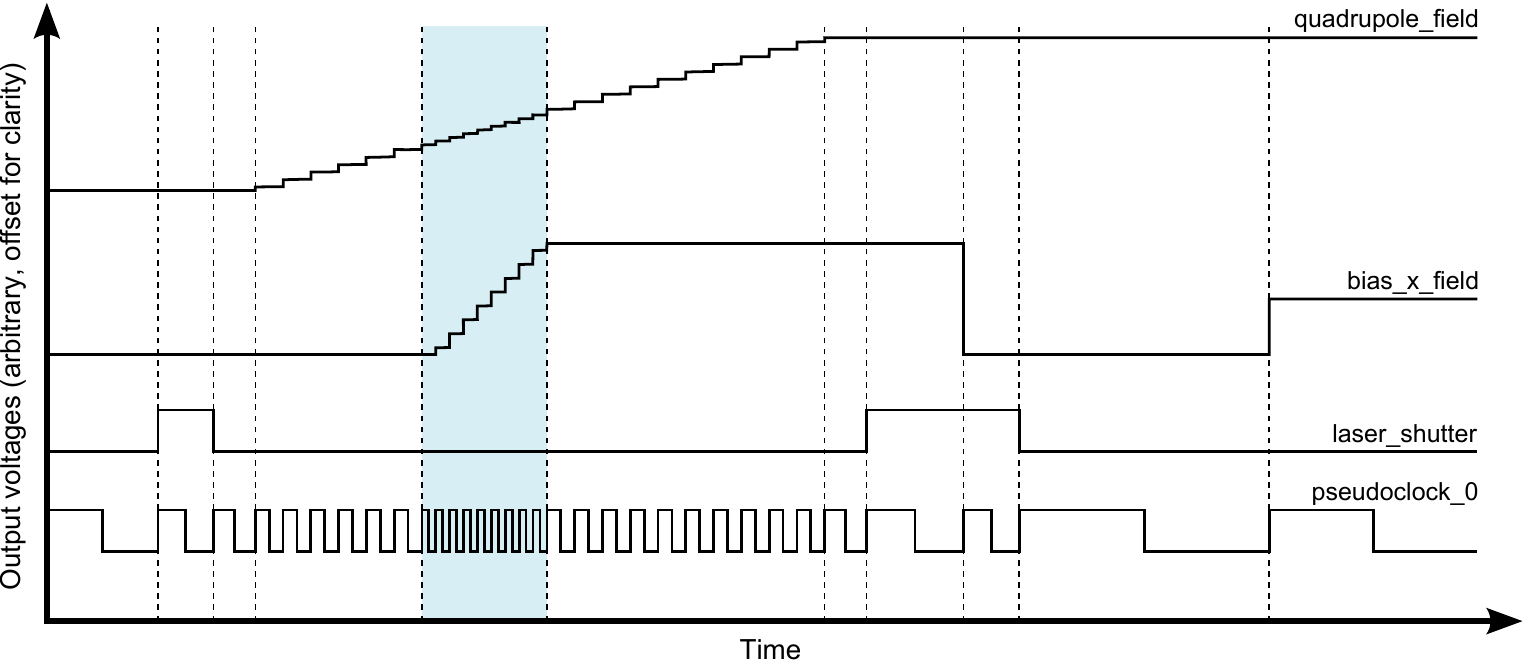}%
\caption{An example of digital and analog voltage outputs generated by the labscript code in Fig. \ref{fig:labscript}.
The pseudoclock (lower trace) ticks when a digital output must change, or at the requested sample rate for time-varying analog outputs (upper two traces).
Dashed vertical lines indicate a change in the pseudoclock frequency.
When multiple analog outputs are varying at the same time (shaded region), the pseudoclock ticks at the highest of their sampling rates.
}%
\label{fig:pseudoclock}%
\end{figure*}

\section{Pseudoclock\label{pseudoclock}}
A typical BEC experiment requires precise timing over a large range of time scales\cite{weidemuller_cold_2009}.
There are periods during which magnetic fields or laser intensities, for example, may change with sub-microsecond resolution.
Conversely, there are periods during which no devices change their output for several seconds, e.g., loading a magneto-optical trap (MOT).
To ensure accurate output during the rapid changes, hardware devices must be preloaded with a set of instructions that can be stepped through by a clock once the experiment begins.
Stepping through instructions at a constant rate requires repetitive instructions during the more inactive periods.
As many devices only support a limited number of instructions, a constant-rate clock limits the maximum sample rate.
A common solution\cite{owen_fast_2004,beeler_matthew_disordered_2011,altin_role_2012,keshet_distributed_2013} is a variable frequency master clock, or \textit{pseudoclock}, which steps through instructions only when a clocked device needs to update an output (see Fig.~\ref{fig:pseudoclock}).
This removes the need for redundant instructions.

All devices sharing a pseudoclock must have an instruction when any one of their outputs changes value.
This can lead to redundant instructions if only some of the devices are changing at a given time.
The instruction limitations of one device may then limit another, e.g., some devices hold only a few thousand instructions in their internal memory, whereas others are limited only by the RAM of the host computer refilling their buffers.
To solve this problem, we employ multiple pseudoclocks, assigning devices of similar memory limitations to the same clock.
At the beginning of a shot, the software starts one pseudoclock (the \textit{master clock}), which then triggers other clocks.

To be a useful pseudoclock, a device must be able to deterministically generate arbitrary digital signals, be hardware triggerable, and hold enough instructions for the required experiment.
We currently use two pseudoclocks: the SpinCore PulseBlaster DDS-II-300-AWG, a commercial device based on a field-programmable gate array (FPGA); and the PineBlaster, a device developed in house based on a microcontroller.
Both devices are externally referenced to a stable 10\,MHz source.

The PineBlaster is a low-cost device using commodity hardware, based on the Arduino-like Digilent ChipKIT Max32 microcontroller prototyping board\footnote{The source code for turning a ChipKIT Max32 into a PineBlaster is available at \protect\url{http://hardware.labscriptsuite.org/}}.
The board is flashed with a program that accepts clock instructions over universal serial bus (USB) and executes them with deterministic timing.
It is capable of clocking at up to 10\,MHz (100\,ns between rising edges) with a resolution of 25\,ns.
The PineBlaster needs one instruction for each change in clock rate (see Fig.~\ref{fig:pseudoclock}) and supports up to 15\,000 instructions.

\texttt{labscript} provides support for adding new pseudoclocks.
It uses an intermediate format for storing timing instructions; implementing a new pseudoclock entails translating them into the required format for the hardware.

Some experiments require the time between instructions to be determined \textit{during} a shot.
This can be achieved by pausing the pseudoclocks until some condition is met.
A common example\cite{keshet_distributed_2013,owen_fast_2004,meyrath_laboratory_2012} is waiting for a sufficient level of fluorescence from a loading MOT.
Both the PulseBlaster and the PineBlaster support \textit{wait instructions}, which pause output until resumed by a trigger.
These instructions, when used in tandem with devices such as voltage comparators, can command the experiment to wait for events of interest.

\begin{figure*}%
\includegraphics{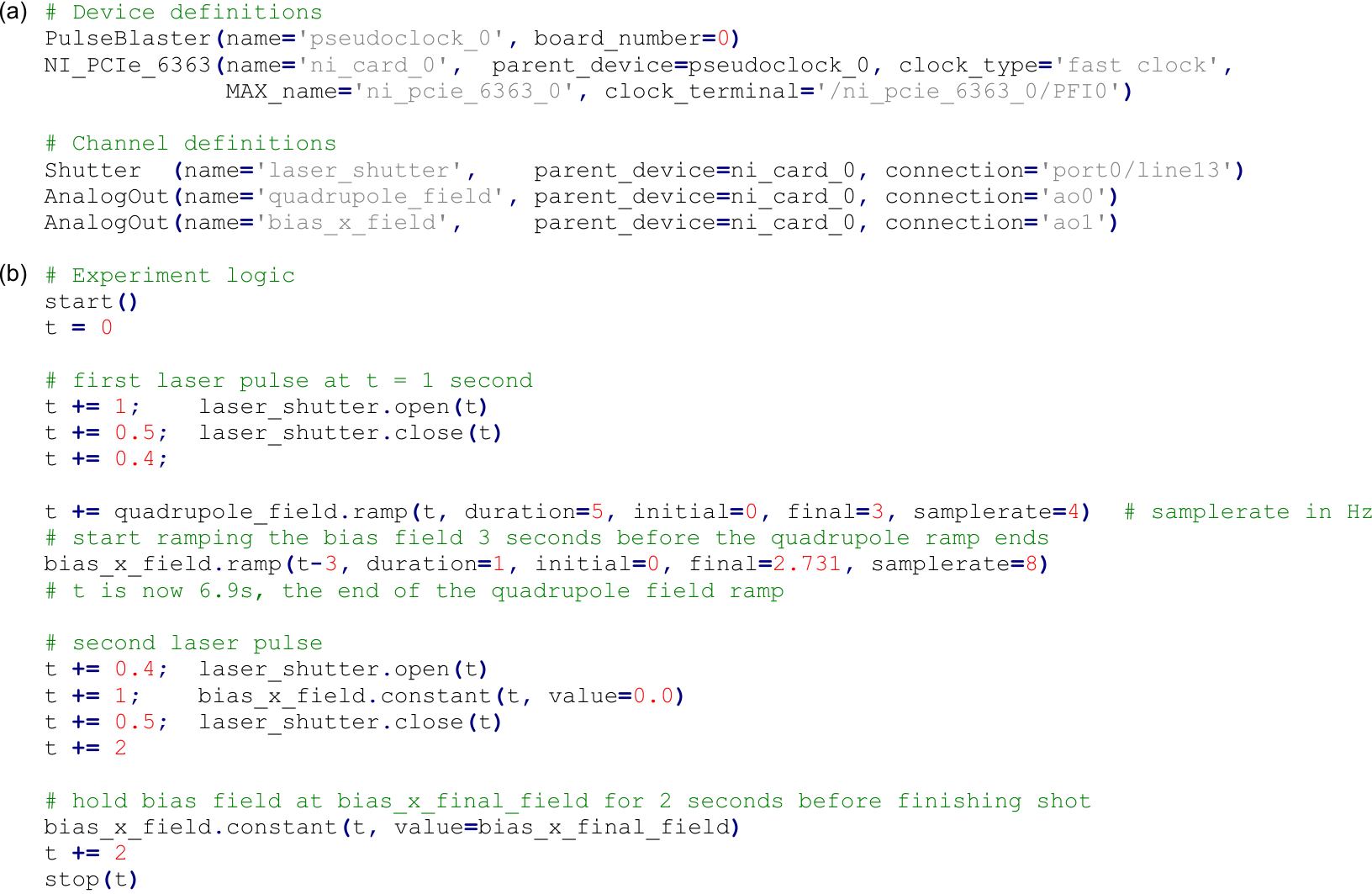}%
\caption{An example \texttt{labscript} file.
The connection table (a) defines a pseudoclock and a multifunction DAC object and configures three output channels.
This is followed by the experiment logic (b) which commands output from these channels by name at times specified by the variable \texttt{t}.
The experiment logic refers to the parameter \texttt{bias\_x\_final\_field} which is set in \texttt{runmanager} (Sec.~\ref{runmanager}).
}%
\label{fig:labscript}%
\end{figure*}

\section{The Labscript Library\label{metalanguage}}
We have created a Python software library, \texttt{labscript}, for defining experiment logic.
\texttt{labscript} provides \textit{hardware abstraction}, a common interface to control heterogeneous hardware.
For example, the \texttt{DigitalOut} class provides \texttt{go\_high(t)} and \texttt{go\_low(t)} functions to set the state of a digital output at time \texttt{t}.
The user calls these functions without regard to the underlying device, its method of programming, or the state of other digital outputs connected to the same device.
Based on an experiment script containing such function calls, \texttt{labscript} automatically generates instructions for output and measurement devices as well as pseudoclocks.
The automatic generation of pseudoclock instructions saves the user from dividing overlapping function ramps into segments (Fig. \ref{fig:pseudoclock}), or manually interpolating output values when a new time point is created on another channel.

An experiment script consists of two parts: a \textit{connection table} (Fig.~\ref{fig:labscript} (a)), and code defining the logic of the experiment (Fig.~\ref{fig:labscript} (b)).
The connection table provides a complete description of devices that are required for the experiment and how they are connected.
\texttt{labscript} creates a set of Python objects based on the connection table, each with associated functions for commanding output or measurement from devices.
The logic of the experiment is then defined by calling these functions with parameters such as time and output value.

\begin{figure*}%
\includegraphics{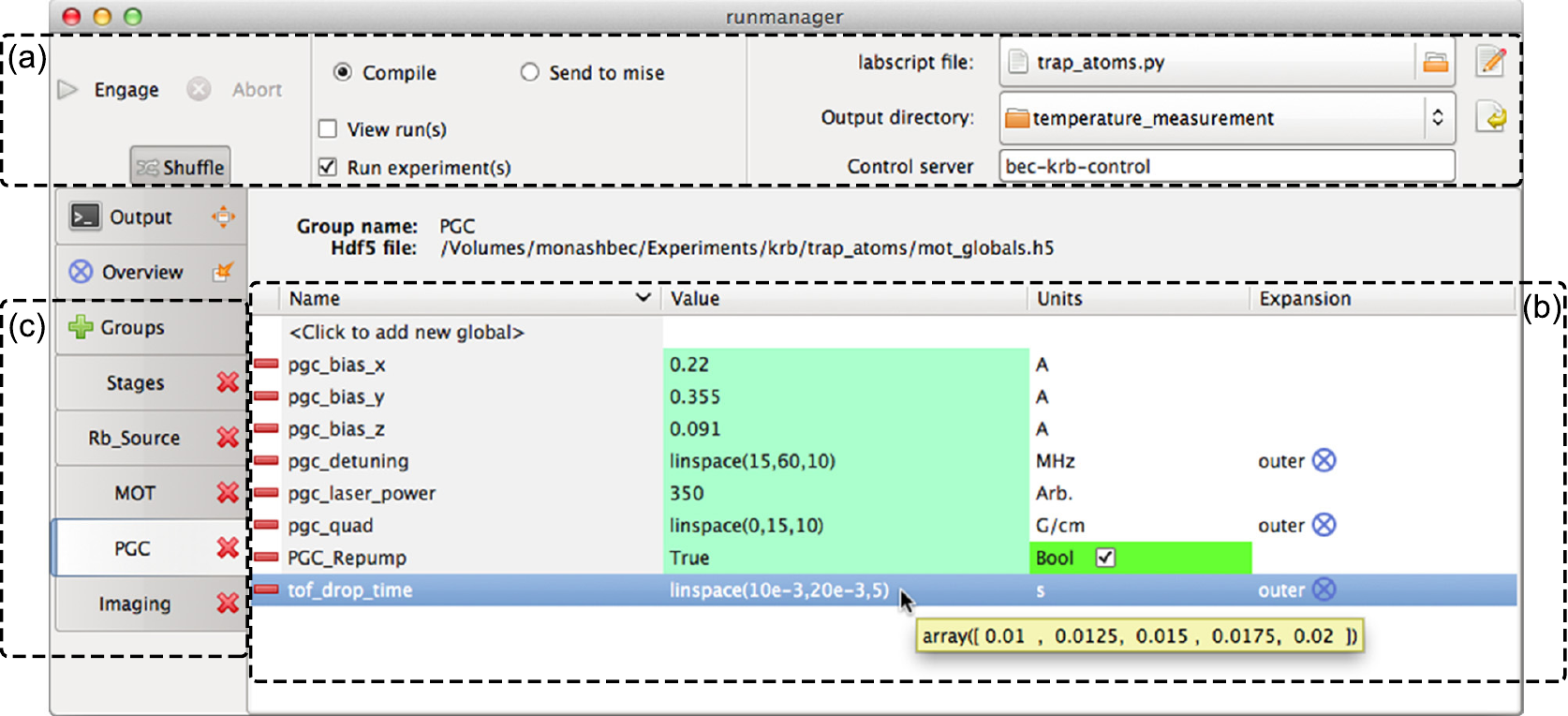}%
\caption{The \texttt{runmanager} interface for configuring experiment parameters.
(a) The experiment logic is specified by the labscript file (here \texttt{trap\_atoms.py}).
HDF files for experiment shots created by \texttt{runmanager} are saved in the output directory.
(b) The value of experiment parameters (``globals'') are specified by Python expressions and may have units.
These can be single values (i.e., \texttt{350} or \texttt{True}), lists, or expressions creating lists (as shown for the globals \texttt{pgc\_detuning}, \texttt{pgc\_quad}, and \texttt{tof\_drop\_time}).
A tooltip shows the evaluation of the global.
The ``Expansion'' column specifies how lists of values are combined to construct a parameter space.
(c) Globals can be separated into groups for convenience.
}%
\label{fig:runmanager}%
\end{figure*}

As the experiment script is executable Python code, the user has full access to standard Python control flow tools, as well as standard and third party Python libraries.
Using a high level language such as Python spares the user from low-level tasks such as memory management \cite{varoquaux_agile_2008}.
User-created functions can be stored in modules and imported into other experiment scripts.
This allows complex experiments to be constructed from simple components, whilst maintaining comprehensibility, resulting in a gentler learning curve for new students.
For example, one might define a \texttt{make\_BEC()} function which contains the logic to form a Bose--Einstein condensate.
While students might not fully understand the experiment logic to create a BEC, they can focus on subsequent experiment logic after a BEC is made.
We have found that text based experiment scripts benefit not just from code re-use but also version control, bug tracking, and comparison of incremental changes (diffs).

When the experiment script is run and a timing sequence created, the \texttt{labscript} functions take into account hardware limitations and provide error messages if these are exceeded.
If no errors are found, the hardware instruction set for all devices in the connection table is written to the HDF file.

While a text-based definition of experiment logic gives a broad overview of the timing sequence, it is not ideal for visualizing the device outputs to ensure the experiment logic is as intended.
The hardware instructions generated by running experiment scripts are difficult to interpret (indeed, \texttt{labscript} was created to mitigate this very problem).
Our program (\texttt{runviewer}) produces plots (similar to Fig.~\ref{fig:pseudoclock}) of the hardware instructions generated by \texttt{labscript}, allowing quick diagnosis of the timing sequence before reaching for the oscilloscope.

\begin{figure*}%
\includegraphics{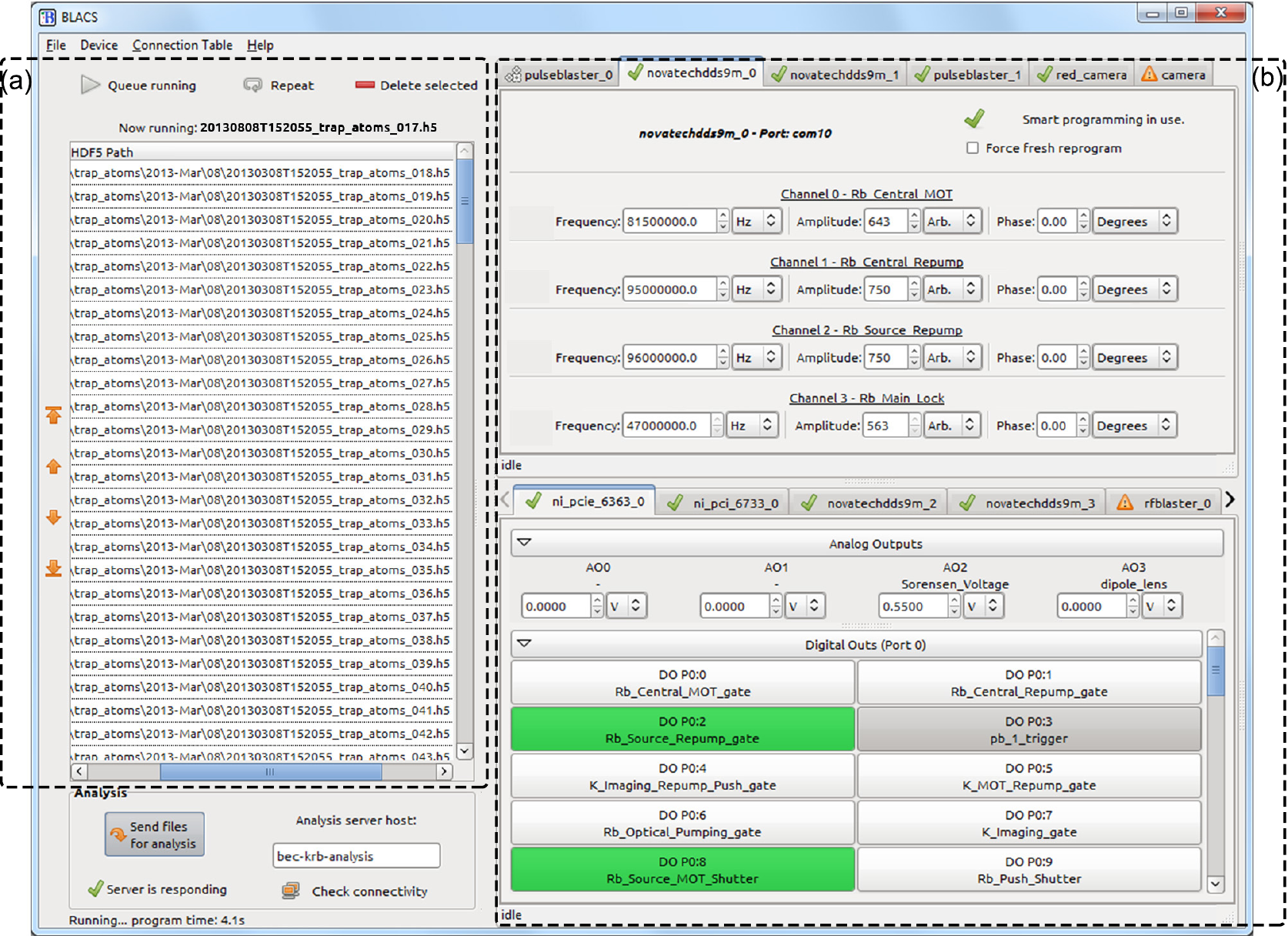}%
\caption{
The \texttt{BLACS} interface for controlling hardware. 
(a) The queue of shots submitted via \texttt{runmanager}.
(b) The manual control interface.
Each tab controls one device.
Controls for all outputs are automatically generated and are named based on the \texttt{BLACS} connection table.
}%
\label{fig:BLACS}%
\end{figure*}

\section{Setting parameters---runmanager}\label{runmanager}

Repeating experiments while varying parameters is a fundamental part of the scientific method.
Anyone who has performed a quantum science experiment will be familiar with tweaking parameters to find a resonance, calibrating a measurement, or acquiring a large amount of scientific data prior to publication.
The logic of the experiment does not change every time a parameter is adjusted, and it is cumbersome to edit numbers in a text file for each modification.

To ameliorate this, \texttt{labscript} experiments can take a series of parameters as input.
The names and values of these parameters are defined in the graphical interface of \texttt{runmanager} (Fig.~\ref{fig:runmanager}).
The values can be any valid Python expression (such as \texttt{0.74}, \texttt{1E-3}, \texttt{sin(pi/2)}, or \texttt{True}) and can refer to each other.
We call these parameters \textit{globals} because they are available as global variables in experiment scripts, where they are simply referred to by name.
For example, these globals might be used to specify the duration of a $\pi$-pulse, the delay between releasing atoms from a trap and imaging them, or the field strengths of bias magnetic coils.
This provides a clean separation between code, which defines the nature of the experiment (such as creating a BEC with a vortex or performing a matter-wave mixing experiment), and parameters that modify individual shots. 

The user may enter a list of values for a global, such as \texttt{[1,2,3]}, or \texttt{linspace(0,10,100)}.
In this case \texttt{runmanager} produces a corresponding list of experiment shots: one for each value. 
If multiple globals are entered as lists, \texttt{runmanager} performs a Cartesian product, creating one shot for each point in the resulting parameter space.
Two or more lists can be \textit{zipped}, in which case \texttt{runmanager} iterates over these lists in lock-step when producing shots.

Specifying globals as lists makes it possible to explore complicated parameter spaces containing hundreds or thousands of shots.
For example, one might investigate how the temperature of laser cooled atoms varies with laser detuning and magnetic field gradient.
Taking the Cartesian product of ten field strengths and ten detunings results in a parameter space of one hundred points.
Thermometry at each point in this parameter space commonly requires multiple shots to characterize the expansion rate of the atom cloud.
A five-shot temperature measurement brings the number of shots to five hundred.
Producing these shots amounts to entering three lists in \texttt{runmanager} and clicking on the ``Engage'' button, as shown in Fig.~\ref{fig:runmanager}.
\texttt{runmanager} then creates five hundred HDF files containing the globals for each shot.
The experiment script is run for each shot, storing hardware instructions in each file\footnote{In our lab, a typical hardware set for running this experiment would be a SpinCore PulseBlaster DDS-II-300-AWG as a pseudoclock, along with a Novatech DDS9m, National Instruments PCIe6363 and PCI6733 boards and a Photonfocus MV1-D1312(I) camera.}.
The HDF files are then submitted to \texttt{BLACS} for execution.

\begin{figure*}%
\includegraphics{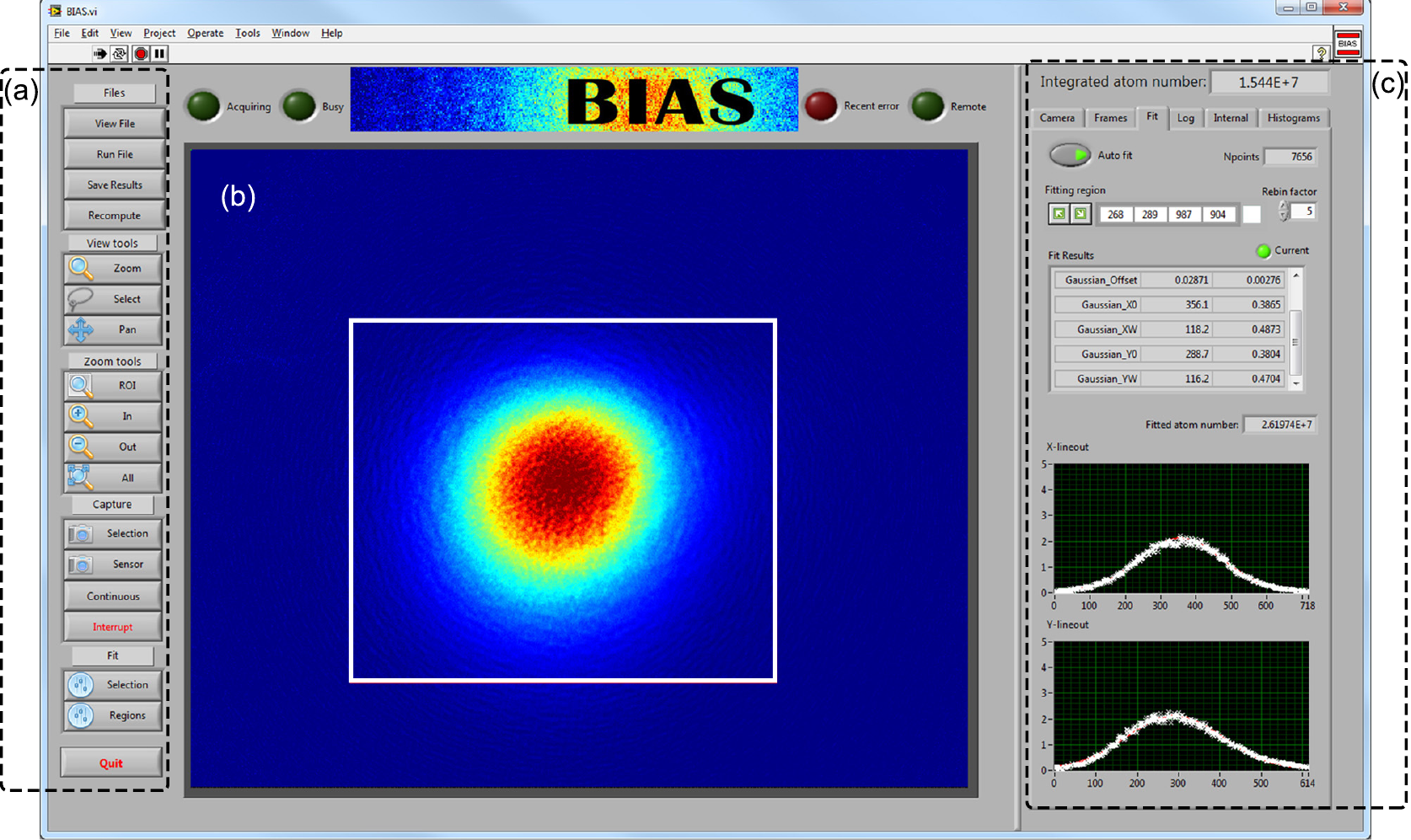}%
\caption{The \texttt{BIAS} interface displaying a laser cooled atom cloud.
(a) Manual controls for loading and capturing images, selecting regions of interest and zooming.
(b) Computed optical depth (OD) image of the atoms, with a region of interest (white) selected for fitting.
Multiple regions of interest may be selected for multi-component atom clouds.
(c) Atom number and cloud size are displayed for immediate feedback.
}
\label{fig:BIAS}%
\end{figure*}

\section{experiment execution---BLACS\label{experiment_execution}}
\texttt{BLACS} coordinates input and output through hardware devices. 
These devices can be local, and thus under the direct control of \texttt{BLACS}, or connected to a different computer as part of a secondary control program such as \texttt{BIAS} (Sec.~\ref{BIAS}).
\texttt{BLACS} provides both manual control of devices (through a GUI) and buffered execution of experiment shots.

The GUI for manual control is dynamically generated from a \textit{lab connection table} that describes the current configuration of all connected devices.
Each device is allocated a tab in the interface, containing controls for commanding output when in manual control mode (Fig.~\ref{fig:BLACS}).

Upon submission to \texttt{BLACS}, HDF files containing hardware instructions are checked for validity and placed in a queue.
The queue can be reordered, paused, or put on repeat.
The validity check compares the connection table of each shot to the lab connection table, rejecting those with incompatible hardware.
This prevents unintended device output that would produce nonsensical results and possibly damage equipment.

\texttt{BLACS} takes the first experiment in the queue, coordinates hardware programming, and sends a start trigger to the master pseudoclock.
The experiment then proceeds under hardware timing.
At the end of a shot, \texttt{BLACS} coordinates saving data acquired by devices to the HDF file, and returns to manual control mode.
Each GUI control is updated to the final values of the shot, maintaining output continuity.

Laboratories are a hostile environment for hardware interface libraries. 
Power cycling of devices and unplugging of cables are common occurrences.
A student tripping over a USB cable (health and safety implications notwithstanding) might be expected to cause an experiment to fail, however the control system ought to recover gracefully when it is plugged back in.
Similarly, bugs in closed source drivers and libraries are points of failure outside of a users control.

To make our system robust against such hardware and software failures, \texttt{BLACS} implements a multiprocess architecture similar to the sandboxed tabs of the Google Chrome web browser\cite{[][{, available at \url{http://seclab.stanford.edu/websec/chromium}}. See \url{http://youtu.be/29e0CtgXZSI} for more information.]barth_security_2008}. 
For each device in \texttt{BLACS}, a \textit{worker process} is spawned, which communicates with the hardware device. 
This makes \texttt{BLACS} robust against crashes: if one device has a problem it will not affect others. 
If a hardware device becomes unresponsive, or the device driver encounters a serious error, its isolation in a separate process prevents the GUI and other devices from suffering the same fate.

Should a worker process crash, the user is presented with the option of restarting the process, which will reload any device libraries it uses. 
It is worth noting that systems implemented in LabVIEW cannot force libraries to reload, so errors leading to an undefined state would only be remedied by restarting the entire control system.

The initialization of hardware in preparation for a shot is an important part of an experiment, and can significantly contribute to the experiment cycle time.
The multiprocess architecture naturally provides for simultaneous programming of hardware devices, resulting in an increased experiment duty cycle.
We have implemented a \textit{smart programming} feature on many of our devices, further decreasing programming time, reprogramming them only if their instructions have changed since the previous shot (on a per-instruction basis when possible).
Devices with large buffers and slow communication (such as the Novatech DDS9m rf synthesizer) benefit greatly from this technique.

\begin{figure*}%
\includegraphics{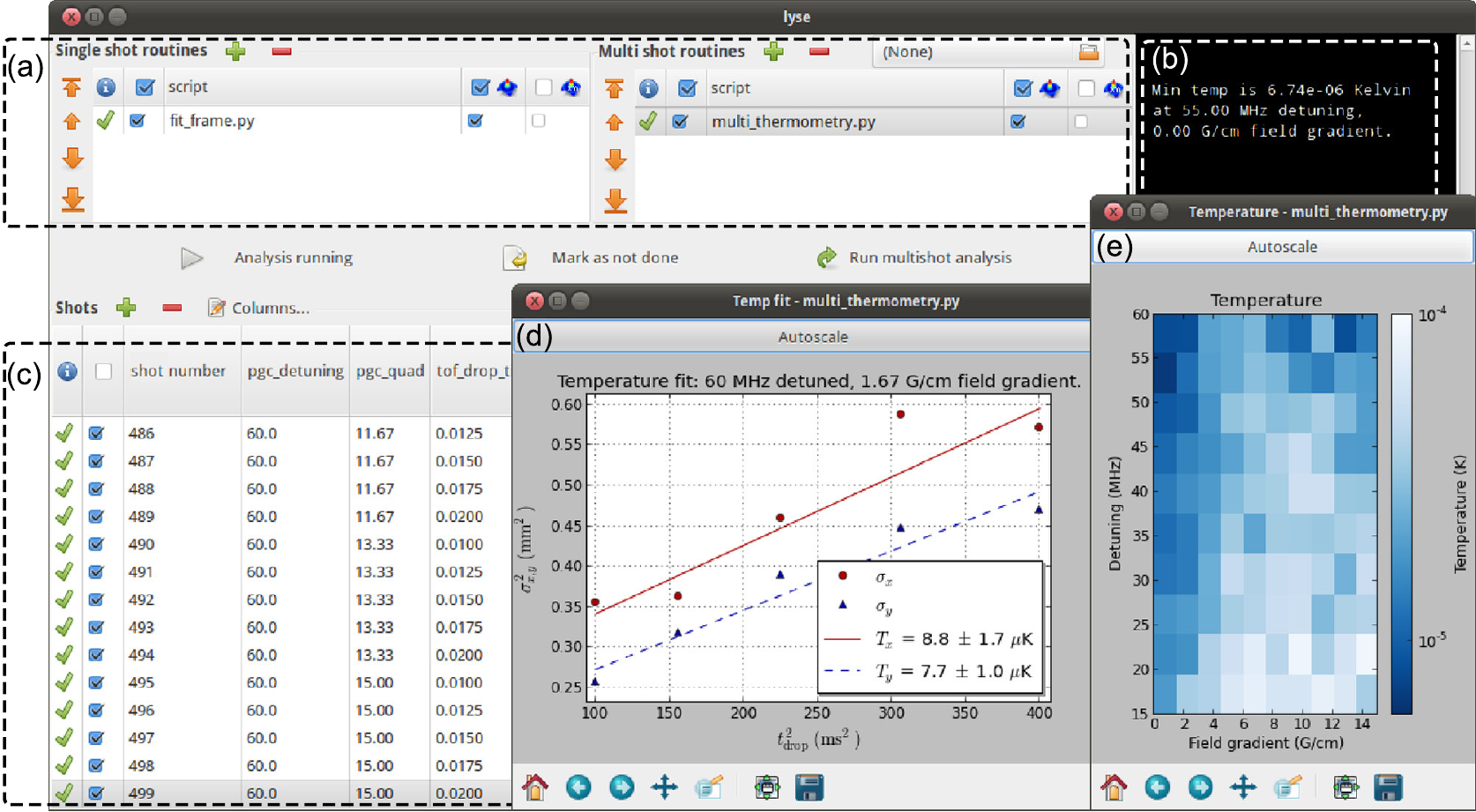}%
\caption{
The \texttt{lyse} interface.
(a) Routines can be selected to analyze single or multiple shots.
(b) Terminal output from the analysis routines in (a).
(c) Table of shots; columns show globals and analysis results.
A small subset of columns is displayed here.
(d) A fit yielding the temperature of laser cooled atoms prepared at a particular field gradient and detuning.
(e) The results of the analysis in (d) repeated at each point in the parameter space.
 }%
\label{fig:lyse}%
\end{figure*}

\section{Image acquisition---BIAS}\label{BIAS}
Using secondary control programs to communicate with specific devices is desirable when software to do so exists and has been debugged, particularly software written in another programming language.
\texttt{BLACS} integrates such programs into the control flow by sending them HDF files containing hardware instructions to program devices for execution upon a hardware trigger.
\texttt{BLACS} notifies secondary control programs that the shot has completed, at which point they write any acquired data to the HDF file.

Our camera control and image acquisition system, \texttt{BIAS}, is one such program.
\texttt{BIAS} is a LabVIEW application that operates scientific cameras, captures image sequences, and performs image processing tasks such as background subtraction, saturation correction, optical depth calculation, and simple 2D fitting.

Multiple instances of \texttt{BIAS} can be run simultaneously to control multiple cameras in one experiment.
\texttt{BIAS} can also run as a stand-alone program for quick visualization of previously captured data or acquire images manually.
Hardware communication in \texttt{BIAS} is abstracted through LabVIEW's object hierarchy, allowing a camera class to be written for any vendor library.

LabVIEW provides convenient components for creating graphical interfaces, and \texttt{BIAS} displays raw and computed images as they become available (Fig.~\ref{fig:BIAS}).
Fit results such as atom cloud shape and atom number are prominently displayed to detect and diagnose problems as they occur.
The camera acquisition area and regions of interest used to inform fits can be interactively adjusted, without needing to interrupt or recreate a currently running sequence of shots.
Multiple regions of interest can be selected and their coordinates saved to the HDF file, enabling further analysis.

\section{Analysis---LYSE}\label{lyse}

Analysis is a critical part of an autonomous control system. 
Automated analysis---performed immediately after every shot---is often restricted to routines that change infrequently and are applied uniformly once per shot.
Ideally analysis should be flexible as well as autonomous; these can be conflicting goals without a unifying analysis framework.
Our analysis system \texttt{lyse} accommodates collective analysis of a group of shots and trivial re-analysis upon changing or adding routines. 

\texttt{lyse} is a scheduler for user-written analysis routines, which are ordinary Python scripts.
It provides functions for extracting the experiment data and metadata from the HDF files and saving analysis results to these files.
Multiple analysis routines added to \texttt{lyse} execute one after the other when a new HDF file is received over the network, or on command through the GUI.
Plots produced by the user's code are updated following every shot as new data comes in from the experiment.

There are two types of routine that \texttt{lyse} can run: single-shot, which are run on every shot, and multi-shot, which analyze a group of shots together.
Analysis of the thermometry example in Sec.~\ref{runmanager} is shown in Fig.~\ref{fig:lyse}.
A single-shot routine computes the size of an atom cloud after a fixed expansion time, and a multi-shot routine uses these results to determine the expansion rate and thus the temperature.
The multi-shot routine then plots this temperature as a function of laser detuning and magnetic field strength.

Splitting, sorting, plotting, and exploring large multidimensional datasets are cumbersome when directly accessing a set of files.
In addition to direct access to the HDF files, \texttt{lyse} provides a tabular data structure---a \texttt{pandas}\cite{mckinney_pandas} DataFrame---for multi-shot routines, containing all globals as set by \texttt{runmanager}, and all single-shot analysis results.
With \texttt{pandas} and the the standard Python scientific stack of \texttt{numpy}\cite{oliphant_numpy}, \texttt{scipy}\cite{jones_scipy_2001}, and \texttt{matplotlib}\cite{[][{; see also ``matplotlib: Python plotting,'' \url{http://matplotlib.org/}.}]hunter_matplotlib:_2007}, \texttt{lyse} provides a powerful environment for analysis \cite{mckinney_python_2012}.

Analysis routines can be run independently of \texttt{lyse} if desired.
This allows the same framework and analysis code to be used for publication preparation.

\begin{figure*}%
\includegraphics{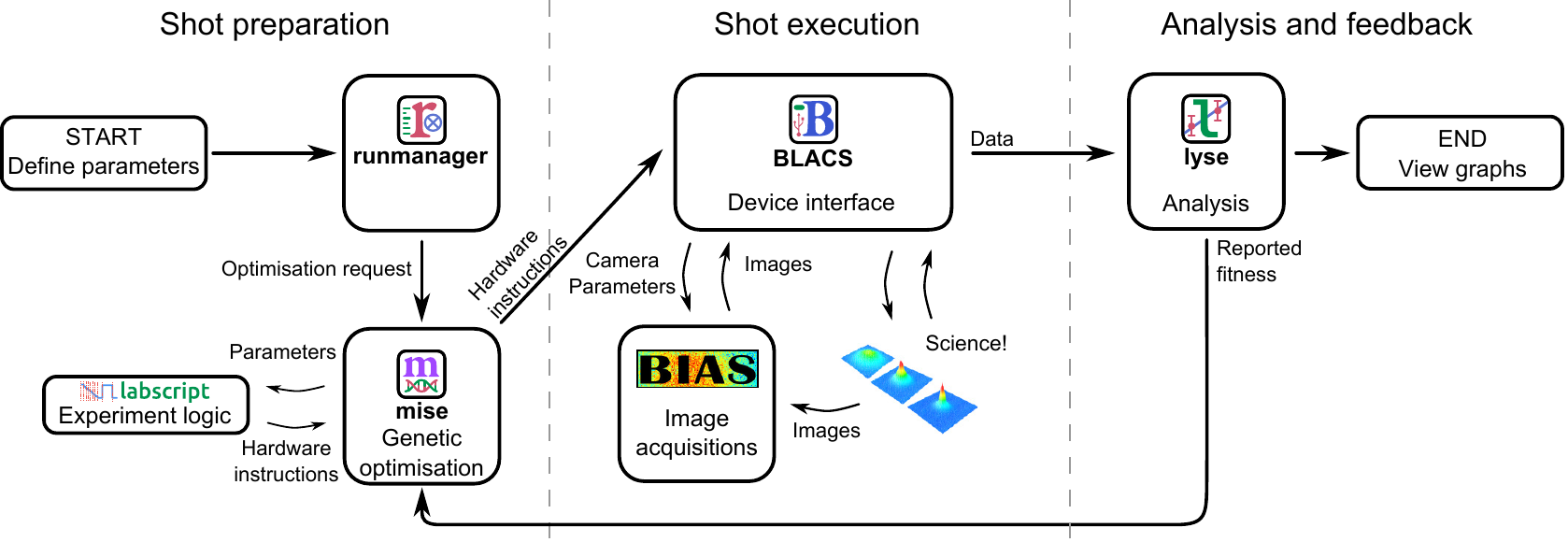}%
\caption{The data flow for closed loop optimization.
In contrast to Fig.~\ref{flow_chart}, analysis results are used to determine future shots automatically.
The optimizer \texttt{mise} varies parameters, directly calling \texttt{labscript} to compile new experiment shots.
Parameters to be optimized are selected by the user in \texttt{runmanager}.
\texttt{lyse} reports fitness to \texttt{mise} which is used to create the next generation of shots.
}%
\label{flow_chart_mise}%
\end{figure*}

\section{Optimization---MISE\label{mise}}

Marrying powerful Python tools to shot-based analysis permits extensibility of the control system, such as closed loop optimization of measured quantities.
One often performs parameter space scans for optimization, requiring many shots.
This may be tuning a parameter of an apparatus to enhance its performance, finding a resonance of some transition, or some other feature of interest.
The quantity being optimized is often the result of some analysis, e.g., the temperature of ultracold atoms (mentioned in Secs.~\ref{runmanager}~and~\ref{lyse}).
We have created \texttt{mise}, a program that performs automatic optimization of analysis results using a genetic algorithm\cite{back_overview_1993}.
A user specifies one or more parameters to optimize against a predefined figure of merit.
Genetic algorithms are resistant to noise, making them particularly useful for optimizing experimental results.

The data flow of the optimization process follows Fig.~\ref{flow_chart_mise}, modifying that shown in Fig.~\ref{flow_chart}.
The user specifies in \texttt{runmanager} one or more parameters to optimize, with upper and lower limits for each.
An analysis routine in \texttt{lyse} reports optimality to \texttt{mise}, which creates shots with modified parameters and submits them to \texttt{BLACS}.

For each parameter being optimized the user also specifies a \textit{mutation rate}.
This determines how much the parameter is varied per generation of the genetic algorithm: the larger the mutation rate, the faster \texttt{mise} will move towards the optimum.
However, a large mutation rate limits the precision to which the optimal parameters can be determined.

With this specification of parameters, \texttt{mise} creates a population of \textit{individuals}.
Each individual comprises values from one point in the optimization parameter space, initially chosen at random. 
An individual may be a single experiment shot, or---when optimizing the result of a multi-shot analysis---a sequence of shots.
Once the shots comprising an individual have executed, the user's analysis routine computes a \textit{fitness}, which may be derived from any measured quantity.
\texttt{mise} uses the reported fitness in the genetic algorithm to optimize the specified parameters.
The genetic algorithm used by \texttt{mise}\footnote{See supplementary material at [URL will be inserted by AIP] for implementation details of the genetic algorithm used by \texttt{mise}} is a variation on pointed directed mutation\cite{berry_pod_2004}, in which mutations are biased in directions previously shown to be successful. 

The user can specify when to stop the optimization, either by manual intervention or by a convergence condition written into their analysis script.
They may also ``guide'' the evolution by adding and deleting individuals from the gene pool at any time.

\begin{figure}%
\includegraphics{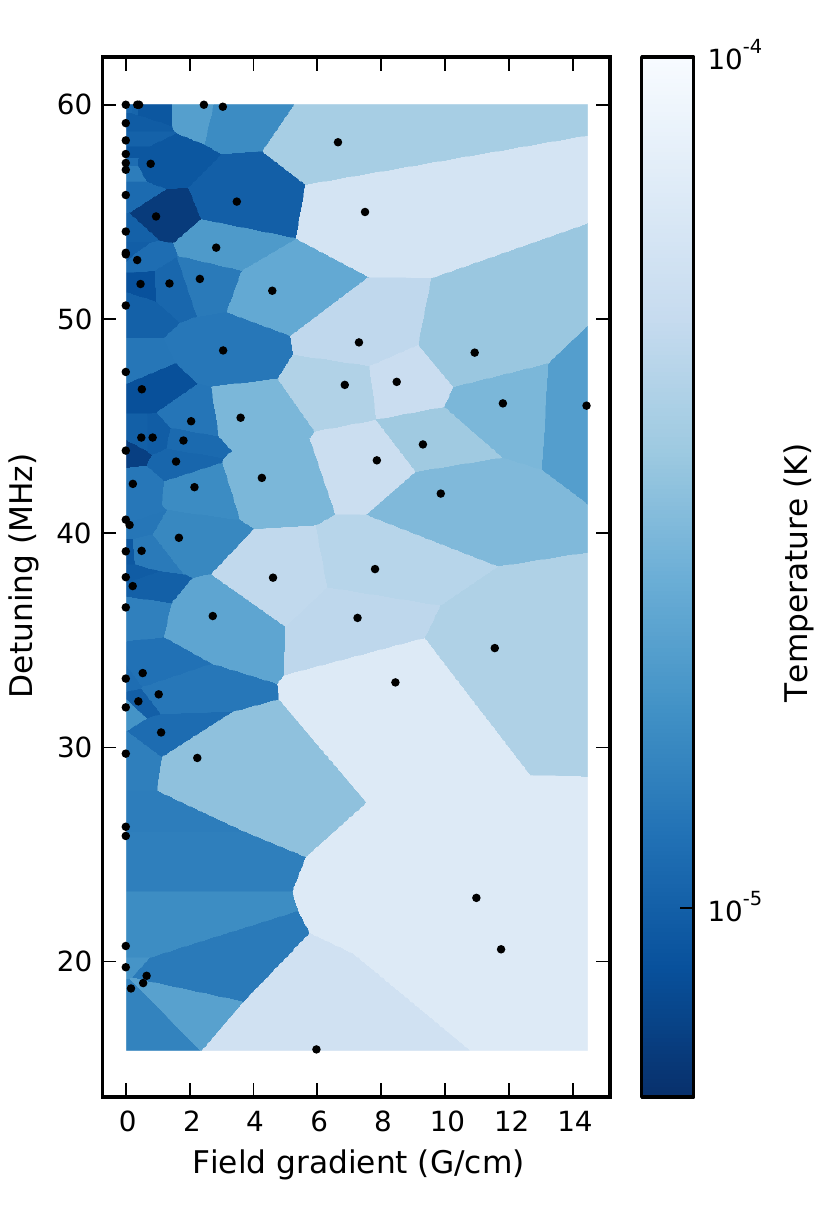}%
\caption{
A proof-of-principle optimization using \texttt{mise}.
\texttt{mise} scanned the parameter space described in Sec.~\ref{runmanager}, searching for the coldest point.
Each black point represents a temperature measurement at a specific field gradient and detuning, with the surrounding shading indicating the temperature.
Eighty points were taken, corresponding to 400 shots.
The colder region of parameter space is sampled more densely than the uniformly-sampled scan shown if Fig.~\ref{fig:lyse}(e), with 500 shots.
}%
\label{fig:mise_result}%
\end{figure}

An example of automated optimization using \texttt{mise} is shown in Fig.~\ref{fig:mise_result}.
By preferentially exploring the more interesting regions of parameter space, autonomous optimization allows optima to be found in fewer shots.

\texttt{mise} uses the \texttt{labscript} software library to create HDF shot files and submit them to \texttt{BLACS}.
Additional user-written components could similarly submit shots to \texttt{BLACS} if more complex programmatic generation of shots is required.

\section{Portability and Extensibility}
Our software runs on Windows, Linux, and OS~X, although \texttt{BLACS} and \texttt{BIAS} compatibility is subject to the availability of appropriate hardware drivers.
If particular devices must be interfaced with a specific computer, operating system, or programming language, a secondary control program (such as \texttt{BIAS}, Sec.~\ref{BIAS}) can be used.
The components of the labscript suite communicate with each other via data in HDF files, and over the network with ZeroMQ sockets.
The widespread support of these technologies across many platforms
\footnote{
  HDF bindings include C/C++, MATLAB, Python, LabVIEW and Mathematica.
  ZeroMQ support includes C/C++, Python, LabVIEW, Java and many more. See \protect\url{http://www.hdfgroup.org/products/hdf5_tools/} and \protect\url{http://www.zeromq.org/bindings:_start/} for more complete lists.
}
ensures users are not bound to any one operating system or programming language.
The modular nature of our system allows users to replace or supplement any of our programs in their choice of language.

The programs themselves are also written with extensibility in mind.
Adding new hardware support to the labscript suite entails writing a new device class for \texttt{labscript}, and a GUI tab for \texttt{BLACS},\footnote{\texttt{BLACS} communicates with hardware devices through user-written interface code. Devices communicating over standard buses (RS232, USB, Ethernet) are easily interfaced using standard Python libraries for these buses. Devices with proprietary interfaces can be programmed by calls to vendor-supplied libraries through Python's sophisticated foreign-function interface.} or a camera class for \texttt{BIAS}.
Adding analysis routines to \texttt{lyse} amounts to writing a Python script to process experiment data.
Existing library functions and base classes assist such development.
The suite has already proved useful in a setting distinct from quantum science experiments, automating the prototyping of an objective lens, in which the image of a pinhole was acquired and analyzed at 3600 points in a plane to determine the field of view\cite{bennie_objective_2013}.

The labscript suite is open-source and freely available online\cite{labscript_web}.
We encourage readers to contact us if they are interested in implementing the suite in their laboratory.

\section*{Acknowledgments}
The authors would like to thank the current users of our system, who were not part of the development team, L. Bennie, M. Egorov, and A. Wood for their input into making this a better system.
This work was supported by Australian Research Council grant Nos. DP1094399 and DP1096830.

\bibliography{main}

\begin{thebibliography}{36}%
\makeatletter
\providecommand \@ifxundefined [1]{%
 \@ifx{#1\undefined}
}%
\providecommand \@ifnum [1]{%
 \ifnum #1\expandafter \@firstoftwo
 \else \expandafter \@secondoftwo
 \fi
}%
\providecommand \@ifx [1]{%
 \ifx #1\expandafter \@firstoftwo
 \else \expandafter \@secondoftwo
 \fi
}%
\providecommand \natexlab [1]{#1}%
\providecommand \enquote  [1]{``#1''}%
\providecommand \bibnamefont  [1]{#1}%
\providecommand \bibfnamefont [1]{#1}%
\providecommand \citenamefont [1]{#1}%
\providecommand \href@noop [0]{\@secondoftwo}%
\providecommand \href [0]{\begingroup \@sanitize@url \@href}%
\providecommand \@href[1]{\@@startlink{#1}\@@href}%
\providecommand \@@href[1]{\endgroup#1\@@endlink}%
\providecommand \@sanitize@url [0]{\catcode `\\12\catcode `\$12\catcode
  `\&12\catcode `\#12\catcode `\^12\catcode `\_12\catcode `\%12\relax}%
\providecommand \@@startlink[1]{}%
\providecommand \@@endlink[0]{}%
\providecommand \url  [0]{\begingroup\@sanitize@url \@url }%
\providecommand \@url [1]{\endgroup\@href {#1}{\urlprefix }}%
\providecommand \urlprefix  [0]{URL }%
\providecommand \Eprint [0]{\href }%
\providecommand \doibase [0]{http://dx.doi.org/}%
\providecommand \selectlanguage [0]{\@gobble}%
\providecommand \bibinfo  [0]{\@secondoftwo}%
\providecommand \bibfield  [0]{\@secondoftwo}%
\providecommand \translation [1]{[#1]}%
\providecommand \BibitemOpen [0]{}%
\providecommand \bibitemStop [0]{}%
\providecommand \bibitemNoStop [0]{.\EOS\space}%
\providecommand \EOS [0]{\spacefactor3000\relax}%
\providecommand \BibitemShut  [1]{\csname bibitem#1\endcsname}%
\let\auto@bib@innerbib\@empty
\bibitem [{\citenamefont {Weidem{\"u}ller}\ and\ \citenamefont
  {Zimmermann}(2009)}]{weidemuller_cold_2009}%
  \BibitemOpen
  \bibfield  {author} {\bibinfo {author} {\bibfnamefont {M.}~\bibnamefont
  {Weidem{\"u}ller}}\ and\ \bibinfo {author} {\bibfnamefont {C.}~\bibnamefont
  {Zimmermann}},\ }\href@noop {} {\emph {\bibinfo {title} {Cold Atoms and
  Molecules}}}\ (\bibinfo  {publisher} {Wiley},\ \bibinfo {year}
  {2009})\BibitemShut {NoStop}%
\bibitem [{\citenamefont {Robins}\ \emph {et~al.}(2013)\citenamefont {Robins},
  \citenamefont {Altin}, \citenamefont {Debs},\ and\ \citenamefont
  {Close}}]{robins_atom_2013}%
  \BibitemOpen
  \bibfield  {author} {\bibinfo {author} {\bibfnamefont {N.}~\bibnamefont
  {Robins}}, \bibinfo {author} {\bibfnamefont {P.}~\bibnamefont {Altin}},
  \bibinfo {author} {\bibfnamefont {J.}~\bibnamefont {Debs}}, \ and\ \bibinfo
  {author} {\bibfnamefont {J.}~\bibnamefont {Close}},\ }\href@noop {}
  {\bibfield  {journal} {\bibinfo  {journal} {Phys. Rep.}\ }\textbf {\bibinfo
  {volume} {529}},\ \bibinfo {pages} {265} (\bibinfo {year}
  {2013})}\BibitemShut {NoStop}%
\bibitem [{\citenamefont {Cronin}, \citenamefont {Schmiedmayer},\ and\
  \citenamefont {Pritchard}(2009)}]{cronin_optics_2009}%
  \BibitemOpen
  \bibfield  {author} {\bibinfo {author} {\bibfnamefont {A.~D.}\ \bibnamefont
  {Cronin}}, \bibinfo {author} {\bibfnamefont {J.}~\bibnamefont
  {Schmiedmayer}}, \ and\ \bibinfo {author} {\bibfnamefont {D.~E.}\
  \bibnamefont {Pritchard}},\ }\href {\doibase 10.1103/RevModPhys.81.1051}
  {\bibfield  {journal} {\bibinfo  {journal} {Rev. Mod. Phys.}\ }\textbf
  {\bibinfo {volume} {81}},\ \bibinfo {pages} {1051} (\bibinfo {year}
  {2009})}\BibitemShut {NoStop}%
\bibitem [{\citenamefont {Negretti}, \citenamefont {Treutlein},\ and\
  \citenamefont {Calarco}(2011)}]{negretti_quantum_2011}%
  \BibitemOpen
  \bibfield  {author} {\bibinfo {author} {\bibfnamefont {A.}~\bibnamefont
  {Negretti}}, \bibinfo {author} {\bibfnamefont {P.}~\bibnamefont {Treutlein}},
  \ and\ \bibinfo {author} {\bibfnamefont {T.}~\bibnamefont {Calarco}},\ }\href
  {\doibase 10.1007/s11128-011-0291-5} {\bibfield  {journal} {\bibinfo
  {journal} {Quantum Inf. Process.}\ }\textbf {\bibinfo {volume} {10}},\
  \bibinfo {pages} {721} (\bibinfo {year} {2011})}\BibitemShut {NoStop}%
\bibitem [{\citenamefont {Ladd}\ \emph {et~al.}(2010)\citenamefont {Ladd},
  \citenamefont {Jelezko}, \citenamefont {Laflamme}, \citenamefont {Nakamura},
  \citenamefont {Monroe},\ and\ \citenamefont
  {{O{\textquoteright}Brien}}}]{ladd_quantum_2010}%
  \BibitemOpen
  \bibfield  {author} {\bibinfo {author} {\bibfnamefont {T.~D.}\ \bibnamefont
  {Ladd}}, \bibinfo {author} {\bibfnamefont {F.}~\bibnamefont {Jelezko}},
  \bibinfo {author} {\bibfnamefont {R.}~\bibnamefont {Laflamme}}, \bibinfo
  {author} {\bibfnamefont {Y.}~\bibnamefont {Nakamura}}, \bibinfo {author}
  {\bibfnamefont {C.}~\bibnamefont {Monroe}}, \ and\ \bibinfo {author}
  {\bibfnamefont {J.~L.}\ \bibnamefont {{O{\textquoteright}Brien}}},\ }\href
  {\doibase 10.1038/nature08812} {\bibfield  {journal} {\bibinfo  {journal}
  {Nature}\ }\textbf {\bibinfo {volume} {464}},\ \bibinfo {pages} {45}
  (\bibinfo {year} {2010})}\BibitemShut {NoStop}%
\bibitem [{\citenamefont {Bloch}, \citenamefont {Dalibard},\ and\ \citenamefont
  {Nascimb{\`e}ne}(2012)}]{bloch_quantum_2012}%
  \BibitemOpen
  \bibfield  {author} {\bibinfo {author} {\bibfnamefont {I.}~\bibnamefont
  {Bloch}}, \bibinfo {author} {\bibfnamefont {J.}~\bibnamefont {Dalibard}}, \
  and\ \bibinfo {author} {\bibfnamefont {S.}~\bibnamefont {Nascimb{\`e}ne}},\
  }\href {\doibase 10.1038/nphys2259} {\bibfield  {journal} {\bibinfo
  {journal} {Nat. Phys.}\ }\textbf {\bibinfo {volume} {8}},\ \bibinfo {pages}
  {267} (\bibinfo {year} {2012})}\BibitemShut {NoStop}%
\bibitem [{\citenamefont {Blatt}\ and\ \citenamefont
  {Roos}(2012)}]{blatt_quantum_2012}%
  \BibitemOpen
  \bibfield  {author} {\bibinfo {author} {\bibfnamefont {R.}~\bibnamefont
  {Blatt}}\ and\ \bibinfo {author} {\bibfnamefont {C.~F.}\ \bibnamefont
  {Roos}},\ }\href {\doibase 10.1038/nphys2252} {\bibfield  {journal} {\bibinfo
   {journal} {Nat. Phys.}\ }\textbf {\bibinfo {volume} {8}},\ \bibinfo {pages}
  {277} (\bibinfo {year} {2012})}\BibitemShut {NoStop}%
\bibitem [{\citenamefont {Varoquaux}(2008)}]{varoquaux_agile_2008}%
  \BibitemOpen
  \bibfield  {author} {\bibinfo {author} {\bibfnamefont {G.}~\bibnamefont
  {Varoquaux}},\ }\href {\doibase 10.1109/MCSE.2008.47} {\bibfield  {journal}
  {\bibinfo  {journal} {Comput. Sci. Eng.}\ }\textbf {\bibinfo {volume} {10}},\
  \bibinfo {pages} {55} (\bibinfo {year} {2008})}\BibitemShut {NoStop}%
\bibitem [{\citenamefont {Gaskell}\ \emph {et~al.}(2009)\citenamefont
  {Gaskell}, \citenamefont {Thorn}, \citenamefont {Alba},\ and\ \citenamefont
  {Steck}}]{gaskell_open-source_2009}%
  \BibitemOpen
  \bibfield  {author} {\bibinfo {author} {\bibfnamefont {P.~E.}\ \bibnamefont
  {Gaskell}}, \bibinfo {author} {\bibfnamefont {J.~J.}\ \bibnamefont {Thorn}},
  \bibinfo {author} {\bibfnamefont {S.}~\bibnamefont {Alba}}, \ and\ \bibinfo
  {author} {\bibfnamefont {D.~A.}\ \bibnamefont {Steck}},\ }\href {\doibase
  doi:10.1063/1.3250825} {\bibfield  {journal} {\bibinfo  {journal} {Rev. Sci.
  Instrum.}\ }\textbf {\bibinfo {volume} {80}},\ \bibinfo {pages} {115103}
  (\bibinfo {year} {2009})}\BibitemShut {NoStop}%
\bibitem [{\citenamefont {Anderson}(2010)}]{anderson_nonequilibrium_2010}%
  \BibitemOpen
  \bibfield  {author} {\bibinfo {author} {\bibfnamefont {R.~P.}\ \bibnamefont
  {Anderson}},\ }\href {http://tiny.cc/RussellAndersonPhD} {\bibinfo {type}
  {{PhD} thesis}},\ \bibinfo  {school} {Swinburne University of Technology}
  (\bibinfo {year} {2010})\BibitemShut {NoStop}%
\bibitem [{\citenamefont {Beeler}(2011)}]{beeler_matthew_disordered_2011}%
  \BibitemOpen
  \bibfield  {author} {\bibinfo {author} {\bibfnamefont {M.}~\bibnamefont
  {Beeler}},\ }\href
  {http://drum.lib.umd.edu/bitstream/1903/11452/1/Beeler_umd_0117E_11919.pdf}
  {\bibinfo {type} {{PhD} thesis}},\ \bibinfo  {school} {University of
  Maryland} (\bibinfo {year} {2011})\BibitemShut {NoStop}%
\bibitem [{\citenamefont {Altin}(2012)}]{altin_role_2012}%
  \BibitemOpen
  \bibfield  {author} {\bibinfo {author} {\bibfnamefont {P.~A.}\ \bibnamefont
  {Altin}},\ }\href {http://atomlaser.anu.edu.au/publications/altin.pdf}
  {\bibinfo {type} {{PhD} thesis}},\ \bibinfo  {school} {Australian National
  University} (\bibinfo {year} {2012})\BibitemShut {NoStop}%
\bibitem [{\citenamefont {St{\"o}ferle}(2005)}]{stoferle_exploring_2005}%
  \BibitemOpen
  \bibfield  {author} {\bibinfo {author} {\bibfnamefont {T.}~\bibnamefont
  {St{\"o}ferle}},\ }\href@noop {} {\bibinfo {type} {{PhD} thesis}},\ \bibinfo
  {school} {Swiss Federal Institute of Technology}, \bibinfo {address} {Zurich}
  (\bibinfo {year} {2005})\BibitemShut {NoStop}%
\bibitem [{\citenamefont {Keshet}\ and\ \citenamefont
  {Ketterle}(2013)}]{keshet_distributed_2013}%
  \BibitemOpen
  \bibfield  {author} {\bibinfo {author} {\bibfnamefont {A.}~\bibnamefont
  {Keshet}}\ and\ \bibinfo {author} {\bibfnamefont {W.}~\bibnamefont
  {Ketterle}},\ }\href {\doibase 10.1063/1.4773536} {\bibfield  {journal}
  {\bibinfo  {journal} {Rev. Sci. Instrum.}\ }\textbf {\bibinfo {volume}
  {84}},\ \bibinfo {eid} {015105} (\bibinfo {year} {2013})}\BibitemShut
  {NoStop}%
\bibitem [{\citenamefont {Owen}\ and\ \citenamefont
  {Hall}(2004)}]{owen_fast_2004}%
  \BibitemOpen
  \bibfield  {author} {\bibinfo {author} {\bibfnamefont {S.~F.}\ \bibnamefont
  {Owen}}\ and\ \bibinfo {author} {\bibfnamefont {D.~S.}\ \bibnamefont
  {Hall}},\ }\href {\doibase 10.1063/1.1630833} {\bibfield  {journal} {\bibinfo
   {journal} {Rev. Sci. Instrum.}\ }\textbf {\bibinfo {volume} {75}},\ \bibinfo
  {pages} {259} (\bibinfo {year} {2004})}\BibitemShut {NoStop}%
\bibitem [{\citenamefont {Meyrath}\ and\ \citenamefont
  {Schreck}(2012)}]{meyrath_laboratory_2012}%
  \BibitemOpen
  \bibfield  {author} {\bibinfo {author} {\bibfnamefont {T.}~\bibnamefont
  {Meyrath}}\ and\ \bibinfo {author} {\bibfnamefont {F.}~\bibnamefont
  {Schreck}},\ }\href@noop {} {\enquote {\bibinfo {title} {{A} laboratory
  control system for cold atom experiments},}\ }\bibinfo {howpublished}
  {\url{http://iqoqi.at/control}} (\bibinfo {year} {2012})\BibitemShut
  {NoStop}%
\bibitem [{\citenamefont {Hintjens}(2013)}]{hintjens_code_2013}%
  \BibitemOpen
  \bibfield  {author} {\bibinfo {author} {\bibfnamefont {P.}~\bibnamefont
  {Hintjens}},\ }\href@noop {} {\emph {\bibinfo {title} {Code Connected Volume
  1: Learning {ZeroMQ}}}}\ (\bibinfo  {publisher} {{CreateSpace} Independent
  Publishing Platform},\ \bibinfo {year} {2013})\BibitemShut {NoStop}%
\bibitem [{\citenamefont {{{The HDF Group}}}(2010)}]{the_hdf_group_hdf5_2012}%
  \BibitemOpen
  \bibfield  {author} {\bibinfo {author} {\bibnamefont {{{The HDF Group}}}},\
  }\href@noop {} {\enquote {\bibinfo {title} {Hierarchical data format version
  5},}\ }\bibinfo {howpublished} {\url{http://www.hdfgroup.org/HDF5}} (\bibinfo
  {year} {2000-2010})\BibitemShut {NoStop}%
\bibitem [{scp()}]{scpi}%
  \BibitemOpen
  \href@noop {} {\enquote {\bibinfo {title} {{IEEE Standard Codes, Formats,
  Protocols, and Commond Commands for Use With IEEE Std 488.1-1987, IEEE
  Standard Digital Interface for Programmable Instrumentation}},}\ }\bibinfo
  {note} {IEEE Std 488.2-1992}\BibitemShut {NoStop}%
\bibitem [{\citenamefont {van Rossum}\ \emph {et~al.}(10  )\citenamefont {van
  Rossum} \emph {et~al.}}]{rossum_python}%
  \BibitemOpen
  \bibfield  {author} {\bibinfo {author} {\bibfnamefont {G.}~\bibnamefont {van
  Rossum}} \emph {et~al.},\ }\href@noop {} {\enquote {\bibinfo {title} {Python
  programming language v2.7},}\ }\bibinfo {howpublished}
  {\url{http://docs.python.org/2.7/}} (\bibinfo {year} {2010--})\BibitemShut
  {NoStop}%
\bibitem [{\citenamefont {Hughes}(2010)}]{hughes_real_2010}%
  \BibitemOpen
  \bibfield  {author} {\bibinfo {author} {\bibfnamefont {J.~M.}\ \bibnamefont
  {Hughes}},\ }\href@noop {} {\emph {\bibinfo {title} {Real World
  Instrumentation with Python: Automated Data Acquisition and Control
  Systems}}}\ (\bibinfo  {publisher} {{O'Reilly} Media, Inc.},\ \bibinfo {year}
  {2010})\BibitemShut {NoStop}%
\bibitem [{Note1()}]{Note1}%
  \BibitemOpen
  \bibinfo {note} {The source code for turning a ChipKIT Max32 into a
  PineBlaster is available at \protect \url
  {http://hardware.labscriptsuite.org/}}\BibitemShut {NoStop}%
\bibitem [{Note2()}]{Note2}%
  \BibitemOpen
  \bibinfo {note} {In our lab, a typical hardware set for running this
  experiment would be a SpinCore PulseBlaster DDS-II-300-AWG as a pseudoclock,
  along with a Novatech DDS9m, National Instruments PCIe6363 and PCI6733 boards
  and a Photonfocus MV1-D1312(I) camera.}\BibitemShut {Stop}%
\bibitem [{\citenamefont {Barth}\ \emph {et~al.}(2008)\citenamefont {Barth},
  \citenamefont {Jackson}, \citenamefont {Reis},\ and\ \citenamefont {{{The
  Google Chrome Team}}}}]{barth_security_2008}%
  \BibitemOpen
  \bibfield  {author} {\bibinfo {author} {\bibfnamefont {A.}~\bibnamefont
  {Barth}}, \bibinfo {author} {\bibfnamefont {C.}~\bibnamefont {Jackson}},
  \bibinfo {author} {\bibfnamefont {C.}~\bibnamefont {Reis}}, \ and\ \bibinfo
  {author} {\bibnamefont {{{The Google Chrome Team}}}},\ }\href@noop {}
  {\enquote {\bibinfo {title} {The security architecture of the chromium
  browser},}\ } (\bibinfo {year} {2008}),\ \bibinfo {note} {technical report,
  Stanford Security Laboratory, 353 Serra Mall, Stanford, California
  94305}\BibitemShut {NoStop}%
\bibitem [{\citenamefont {McKinney}()}]{mckinney_pandas}%
  \BibitemOpen
  \bibfield  {author} {\bibinfo {author} {\bibfnamefont {W.}~\bibnamefont
  {McKinney}},\ }\href@noop {} {\enquote {\bibinfo {title} {{pandas: a Python
  data analysis library}},}\ }\bibinfo {howpublished}
  {\url{http://pandas.pydata.org/}}\BibitemShut {NoStop}%
\bibitem [{\citenamefont {Oliphant}()}]{oliphant_numpy}%
  \BibitemOpen
  \bibfield  {author} {\bibinfo {author} {\bibfnamefont {T.}~\bibnamefont
  {Oliphant}},\ }\href@noop {} {\enquote {\bibinfo {title} {{NumPy: numerical
  Python}},}\ }\bibinfo {howpublished}
  {\url{http://www.numpy.org/}}\BibitemShut {NoStop}%
\bibitem [{\citenamefont {Jones}\ \emph {et~al.}(01  )\citenamefont {Jones},
  \citenamefont {Oliphant}, \citenamefont {Peterson} \emph
  {et~al.}}]{jones_scipy_2001}%
  \BibitemOpen
  \bibfield  {author} {\bibinfo {author} {\bibfnamefont {E.}~\bibnamefont
  {Jones}}, \bibinfo {author} {\bibfnamefont {T.}~\bibnamefont {Oliphant}},
  \bibinfo {author} {\bibfnamefont {P.}~\bibnamefont {Peterson}},  \emph
  {et~al.},\ }\href@noop {} {\enquote {\bibinfo {title} {{SciPy}: open source
  scientific tools for {Python}},}\ }\bibinfo {howpublished}
  {\url{http://www.scipy.org/}} (\bibinfo {year} {2001--})\BibitemShut
  {NoStop}%
\bibitem [{\citenamefont {Hunter}(2007)}]{hunter_matplotlib:_2007}%
  \BibitemOpen
  \bibfield  {author} {\bibinfo {author} {\bibfnamefont {J.}~\bibnamefont
  {Hunter}},\ }\href {\doibase 10.1109/MCSE.2007.55} {\bibfield  {journal}
  {\bibinfo  {journal} {Comput. Sci. Eng.}\ }\textbf {\bibinfo {volume} {9}},\
  \bibinfo {pages} {90 } (\bibinfo {year} {2007})}\BibitemShut {NoStop}%
\bibitem [{\citenamefont {{McKinney}}(2012)}]{mckinney_python_2012}%
  \BibitemOpen
  \bibfield  {author} {\bibinfo {author} {\bibfnamefont {W.}~\bibnamefont
  {{McKinney}}},\ }\href@noop {} {\emph {\bibinfo {title} {Python for Data
  Analysis}}}\ (\bibinfo  {publisher} {{O'Reilly} Media, Inc.},\ \bibinfo
  {year} {2012})\BibitemShut {NoStop}%
\bibitem [{\citenamefont {B{\"a}ck}\ and\ \citenamefont
  {Schwefel}(1993)}]{back_overview_1993}%
  \BibitemOpen
  \bibfield  {author} {\bibinfo {author} {\bibfnamefont {T.}~\bibnamefont
  {B{\"a}ck}}\ and\ \bibinfo {author} {\bibfnamefont {H.-P.}\ \bibnamefont
  {Schwefel}},\ }\href {\doibase 10.1162/evco.1993.1.1.1} {\bibfield  {journal}
  {\bibinfo  {journal} {Evol. Comput.}\ }\textbf {\bibinfo {volume} {1}},\
  \bibinfo {pages} {1{\textendash}23} (\bibinfo {year} {1993})}\BibitemShut
  {NoStop}%
\bibitem [{Note3()}]{Note3}%
  \BibitemOpen
  \bibinfo {note} {See supplementary material at [URL will be inserted by AIP]
  for implementation details of the genetic algorithm used by \protect \texttt
  {mise}}\BibitemShut {NoStop}%
\bibitem [{\citenamefont {Berry}\ and\ \citenamefont
  {Vamplew}(2004)}]{berry_pod_2004}%
  \BibitemOpen
  \bibfield  {author} {\bibinfo {author} {\bibfnamefont {A.}~\bibnamefont
  {Berry}}\ and\ \bibinfo {author} {\bibfnamefont {P.}~\bibnamefont
  {Vamplew}},\ }in\ \href@noop {} {\emph {\bibinfo {booktitle} {{AISAT} 2004:
  The 2nd International Conference on Artificial Intelligence in Science and
  Technology}}}\ (\bibinfo {address} {Hobart, Tasmania, Australia},\ \bibinfo
  {year} {2004})\ pp.\ \bibinfo {pages} {200--205}\BibitemShut {NoStop}%
\bibitem [{Note4()}]{Note4}%
  \BibitemOpen
  \bibinfo {note} {HDF bindings include C/C++, MATLAB, Python, LabVIEW and
  Mathematica. ZeroMQ support includes C/C++, Python, LabVIEW, Java and many
  more. See \protect \url {http://www.hdfgroup.org/products/hdf5_tools/} and
  \protect \url {http://www.zeromq.org/bindings:_start/} for more complete
  lists.}\BibitemShut {Stop}%
\bibitem [{Note5()}]{Note5}%
  \BibitemOpen
  \bibinfo {note} {\protect \texttt {BLACS} communicates with hardware devices
  through user-written interface code. Devices communicating over standard
  buses (RS232, USB, Ethernet) are easily interfaced using standard Python
  libraries for these buses. Devices with proprietary interfaces can be
  programmed by calls to vendor-supplied libraries through Python's
  sophisticated foreign-function interface.}\BibitemShut {Stop}%
\bibitem [{\citenamefont {Bennie}\ \emph {et~al.}(2013)\citenamefont {Bennie},
  \citenamefont {Starkey}, \citenamefont {Jasperse}, \citenamefont
  {Billington}, \citenamefont {Anderson},\ and\ \citenamefont
  {Turner}}]{bennie_objective_2013}%
  \BibitemOpen
  \bibfield  {author} {\bibinfo {author} {\bibfnamefont {L.~M.}\ \bibnamefont
  {Bennie}}, \bibinfo {author} {\bibfnamefont {P.~T.}\ \bibnamefont {Starkey}},
  \bibinfo {author} {\bibfnamefont {M.}~\bibnamefont {Jasperse}}, \bibinfo
  {author} {\bibfnamefont {C.~J.}\ \bibnamefont {Billington}}, \bibinfo
  {author} {\bibfnamefont {R.~P.}\ \bibnamefont {Anderson}}, \ and\ \bibinfo
  {author} {\bibfnamefont {L.~D.}\ \bibnamefont {Turner}},\ }\href {\doibase
  10.1364/OE.21.009011} {\bibfield  {journal} {\bibinfo  {journal} {Opt.
  Express}\ }\textbf {\bibinfo {volume} {21}},\ \bibinfo {pages} {9011}
  (\bibinfo {year} {2013})}\BibitemShut {NoStop}%
\bibitem [{lab()}]{labscript_web}%
  \BibitemOpen
  \href@noop {} {\enquote {\bibinfo {title} {The labscript suite: an open
  source experiment control and analysis system},}\ }\bibinfo {howpublished}
  {\url{http://labscriptsuite.org/}}\BibitemShut {NoStop}%
\end{thebibliography}%
\end{document}